\documentclass[a4paper]{article}
\usepackage{amssymb}
\usepackage{amsfonts}
\usepackage{amsmath}
\usepackage{graphicx}
\usepackage{hyperref}
\hypersetup{backref,
            pagebackref=true,
            ps2pdf,
            bookmarks=true,
            bookmarksnumbered=true,
            pdfauthor=,
            colorlinks=true,
            linkcolor=blue}

\title{The Boundary Conformal Field Theories of the 2D Ising critical points}

\author{Smain BALASKA\thanks{%
sbalaska@yahoo.com} and\textit{\ }Nahed Sihem BOUNOUA  \thanks{%
bounoua\_sihem@yahoo.fr\ \ } \\
\textit{Laboratoire de Physique Th\'{e}orique d'Oran. }\\
\textit{D\'{e}partement de Physique. Universit\'{e} d'Oran.}\\
\textit{BP 1524 El'Manouer 31000 Oran. Alg\'{e}ria.}}

\begin{document}

\maketitle

\begin{abstract}
We present a new method to identify the Boundary Conformal Field Theories (BCFTs) describing the critical points
of the Ising model on the strip. It consists in measuring the low-lying excitation energies spectra of its quantum
spin chain for different boundary conditions and then to
compare them with those of the different boundary conformal field theories of the $(A_2,A_3)$ minimal model 
\end{abstract}

\section{Introduction}
The critical properties of the two-dimensional (2D) Ising model on
the plane are known to be described by the $(A_{2},A_{3})$
conformal model, which is a unitary minimal conformal field theory
of the ADE classification. On two dimensional geometries with
boundaries they should be described by conformal field theories
(CFTs) defined on the half plane or equivalently on
the strip (for review see \cite{car1} \cite{z1}, \cite{z2},\cite{z3},\cite%
{z4} and \cite{z5}). \ For each boundary conditions, there exists
a boundary conformal field theory (BCFT) having a spectrum formed
by one or more of the Verma modules of the corresponding CFT on
the plane. The main object of this work is a numerical study of
the Ising linear quantum chain at criticality to identify the
BCFTs of the $(A_{2},A_{3})$ conformal model. \\ The paper is
organized as follows. In the first section we introduce briefly
the subject of boundary conformal field theories and give their
predictions (boundary states, partition functions, operators
contents and energy spectrum) concerning the $(A_{2},A_{3})$
conformal model. In section two we introduce the methods used to
make the identification of the different BCFTs for the 2D Ising
model with the different allowed boundary conditions. The last
section is consacred for presenting our numerical data for the
quantum linear chain realizations of the Ising singularity of the
2D Ising model with different boundary conditions and to identify
them with BCFTs candidates of $(A_{2},A_{3})$ conformal model.
\section{BCFT predictions}
A conformal field theory constructed on the complex upper half
plane will be strongly constrained by the presence of \ the real
axe. In this case only symmetries preserving this boundary are
concerned. So that, the independence of the holomorphic and the
anti-holomorphic components of the Virasoro algebra is broken and
only the holomorphic symmetries for example are considered. As a
consequence, the partition function of a BCFT on the half plane is
no more a sesquilinear combination of the Virasoro characters, as
it was the case in the plane geometry. \ \ The partition functions
$\mathcal{Z}_{\alpha \mathcal{j}\beta }$ describing the evolution
of the system between the boundary states $\left\vert \alpha
\right\rangle $ and $\mathcal{j}\beta \rangle $ become linear on
the characters, as follows (\cite{car1}, \cite{z1}):
\begin{equation}
\mathcal{Z}_{\alpha \mathcal{j}\beta }(q)=\sum_{j}n_{j\alpha }^{\rm{ ~}%
\beta }\chi _{j}(q)  \label{partion function1}
\end{equation}%
in this equation, the indices $j$ runs over all the characters
$\chi _{j}(q)$ of the Virasoro algebra and the multiplicities
$n_{j\alpha }^{\rm{ \quad ~}\beta }$ are positive integers giving
the operator content of the BCFT. They are obtained from the
Verlinde formula \cite{ver} :

\begin{equation}
n_{j\alpha }^{\rm{ \quad ~}\beta }=\sum_{i}\frac{S_{ji}S_{\alpha
i}^{\ast }S_{\beta i}}{S_{1i}}  \label{18}
\end{equation}%
The allowed physical boundary states are some combinations of
Ishibashi states \cite{ishi} of the following form :

\begin{equation}
\left\vert \alpha \right\rangle =\sum_{j}\frac{S_{\alpha j}}{\sqrt{S_{1j}}}%
\mathcal{j}j\rangle \rangle  \label{17}
\end{equation}%
For the $(A_{2},A_{3})$ conformal model we have three representations
generated by the three primary fields
\[
\phi _{(1,1)}=I\qquad \rm{,}\qquad \phi _{(1,2)}=\sigma \ \ \ \ \ \rm{%
and }\ \ \ \ \ \ \ \ \ \phi _{(2,1)}=\epsilon \
\]%
Their conformal weights are respectively $h_{1,1}=0$ , $h_{1,2}=\frac{1}{16}$
and $h_{2,1}=\frac{1}{2}$. In this case the modular matrix is
\[
S=\frac{1}{2}\left(
\begin{array}{ccc}
1 & 1 & \sqrt{2} \\
1 & 1 & -\sqrt{2} \\
\sqrt{2} & -\sqrt{2} & 0%
\end{array}%
\right)
\]
Applying the relation (\ref{17}) we obtain the three allowed boundary states
\begin{eqnarray*}
\left\vert I\right\rangle &=&\frac{1}{\sqrt{2}}\mathcal{j}I\rangle \rangle +%
\frac{1}{\sqrt{2}}\mathcal{j}\epsilon \rangle \rangle +\frac{1}{(2)^{\frac{1%
}{4}}}\mathcal{j}\sigma \rangle \rangle \\
\left\vert \epsilon \right\rangle &=&\frac{1}{\sqrt{2}}\mathcal{j}I\rangle
\rangle +\frac{1}{\sqrt{2}}\mathcal{j}\epsilon \rangle \rangle -\frac{1}{%
(2)^{\frac{1}{4}}}\mathcal{j}\sigma \rangle \rangle \\
\left\vert \sigma \right\rangle &=&\mathcal{j}I\rangle \rangle -\mathcal{j}%
\epsilon \rangle \rangle
\end{eqnarray*}
From (\ref{partion function1}) and (\ref{18}) we obtain the
allowed partition functions
\begin{eqnarray*}
\mathcal{Z}_{\sigma |\sigma } &=&\chi _{(1,1)}+\chi _{(2,1)}=\chi _{I}+\chi
_{\epsilon } \\
\mathcal{Z}_{I|I} &=&\mathcal{Z}_{\epsilon |\epsilon }=\chi _{(1,1)}=\chi
_{I} \\
\mathcal{Z}_{\sigma |I} &=&\mathcal{Z}_{\sigma |\epsilon }=\chi
_{(1,2)}=\chi _{\sigma } \\
\mathcal{Z}_{\epsilon |I} &=&\mathcal{Z}_{I|\epsilon }=\chi _{(2,1)}=\chi
_{\epsilon }
\end{eqnarray*}
The normalized Low-Lying excitation spectra of these four CFT
candidates are given in table(\ref{tablebcftsising})
\begin{center}
\begin{table}[h] \centering%
\begin{tabular}{|l|l|l|l|l|l|l|l|l|}
\hline
\multicolumn{9}{|l|}{\textbf{CFT of }$h_{1,1}$ \textbf{verma module}} \\
\hline excitation energy & 2 & 3 & 4 & 5 & 6 & 7 & 8 & 9 \\ \hline
degeneracy & 1 & 1 & 2 & 2 & 3 & 3 & 5 & 5 \\ \hline
\multicolumn{9}{|l|}{} \\ \hline
\multicolumn{9}{|l|}{\textbf{CFT of \ }$h_{1,2}$\textbf{\ verma module}} \\
\hline excitation energy & 1 & 2 & 3 & 4 & 5 & 6 & 7 & 8 \\ \hline
degeneracy & 1 & 1 & 2 & 2 & 3 & 4 & 5 & 6 \\ \hline
\multicolumn{9}{|l|}{} \\ \hline
\multicolumn{9}{|l|}{\textbf{CFT of \ }$h_{2,1}$\textbf{\ verma module}} \\
\hline excitation energy & 1 & 2 & 3 & 4 & 5 & 6 & 7 & 8 \\ \hline
degeneracy & 1 & 1 & 1 & 2 & 2 & 3 & 4 & 5 \\ \hline
\multicolumn{9}{|l|}{} \\ \hline \multicolumn{9}{|l|}{\textbf{CFT
of \ (}$h_{1,1}\oplus $\textbf{\ }$h_{2,1})$ \textbf{verma
module}}
\\ \hline excitation energy & 1 & 3 & 4 & 5 & 6 & 7 & 8 & 9 \\
\hline degeneracy & 1 & 1 & 1 & 1 & 1 & 1 & 2 & 2
\\ \hline
\end{tabular}%
\caption{excitation spectra of the BCFTs candidates in the$(A_{2},A_{3})$minimal model}\label{tablebcftsising}%
\end{table}%
\end{center}
\section{The Methods}
The Hamiltonian limit of the statistical 2D Ising model provides a
quantum spin chain model having the same critical behavior. We
will focus on the Ising singularity of such quantum spin chain
model including special boundary magnetic fields at the first and
the last sites. \ That is, different boundary conditions are
imposed on the quantum spin chain. We will determine its low lying
excitation spectrum for different chain lengths and at special
values of the spin-spin coupling constant which we will call
"pseudo-critical" values. The phenomenological renormalization
group (PRG) will fixe these special values that define the Ising
singularity (see \cite{henkel} and the references therein).
\\For a given boundary condition, the numerical measurements of
the spectrum for different chain lengths lead to series of values.
When fitted to leading scaling behavior, these values will
correspond to the energy levels at criticality.
\\In the last step, this will be compared with the
BCFT predictions given in table(\ref{tablebcftsising}).\\
The general form of the hamiltonian describing the quantum Ising
chain with magnetic fields at the first and the last sites is
given by
\begin{equation}
H=-\sum_{n=1}^{N-1}t\sigma _{z}(n)\sigma _{z}(n+1)-\sum_{n=1}^{N}\sigma
_{x}(n)-h_{1}\sigma _{z}(1)-h_{2}\sigma _{z}(N)  \label{Isingquantik}
\end{equation}
In equation (\ref{Isingquantik}), $N$ represents the number of
sites in the chain, $\sigma _{x}(n)$ and $\sigma _{z}(n)$ are the
2x2 Pauli spin matrices at site "$n$", $h_{1}$\ and $h_{2}$ are
respectively the external magnetic fields applied at the first and
the last sites and "$t$" is the ferromagnetic spin-spin coupling.
We distinguish four cases:
\begin{enumerate}
\item The free-free boundary conditions. In this case, one sets $%
h_{1}=h_{2}=0$.

\item The fixed-parallel boundary conditions. In this case
one sets:$h_{1}=h_{2}=h_{0}$.

\item The fixed anti-parallel boundary conditions. In this cases one sets, $h_{1}=-h_{2}=h_{0}$

\item The free-fixed boundary conditions. In this
cases one sets:$h_{1}=0$ and $h_{2}=h_{0}$
\end{enumerate}
At the Ising singularity, the special coupling constants solve the PRG
equation \cite{Zuber}
\begin{equation}
\lbrack N-1]m(t(N),N-1)-[N]m(t(N),N)=0
\end{equation}%
where $m(t,N)$ is the energy gap $\left[
E_{1}(t,N)-E_{0}(t,N)\right] $. The energies $E_{0}(t,N)$ and
$E_{1}(t,N)$ are the energies of the respective ground state "0"
and first excited state "1" for the Ising quantum spin chain of
length $N$. As $N\rightarrow \infty $, the special values of $t$
solving the PRG equation, converge to the Ising singularity. The
scaling behavior of physical quantities at solutions of the PRG
equation provide the scaling behavior of the same physical
quantities near the corresponding critical point.
\section{Numerical Check of the BCFT Predictions}
For all the boundary conditions, we have measured the energy
spectrum of the Ising quantum chains of lengths  $N=6$ to $N=12$.
The study of these spectra for different values of $N$ permits to
us to obtain the series describing the energy levels. Then we have
extrapolated them for $N\rightarrow \infty$ to obtain the energy
spectra at criticality. \\
For both the free-free and fixed-parallel boundary conditions the
solutions $t_c(N)$ of the PRG equation are presented respectively
in tables (\ref{table 2}) and (\ref{tcfixfix}). They converge
toward the critical value $t_{c}(\infty)$ $=1$ when
($N\longrightarrow \infty $).

\begin{center}
\begin{table}[ht]
\begin{tabular}{l|l|l|l|l|l|l|l|l|l|l|l}
\hline $N=$ & 4 & 5 & 6 & 7 & 8 & 9 &
10 & 11 & 12\\
\hline  $t_c$ & {\tiny 1.09269}& {\tiny 1.05685 } & {\tiny
1.03846} & {\tiny 1.02775} & {\tiny 1.02097} & {\tiny
1.01640} & {\tiny 1.01318} & {\tiny 1.01083} & {\tiny 1.00905} \\
\hline
\end{tabular}%
\caption{\emph{The critical values of t solving the PRG equation
for free-free B.Cs}}\label{table 2}
\end{table}
\end{center}
\begin{center}
\begin{table}[h] \centering%
\begin{tabular}{l|l|l|l|l|l|l|l|l|l|l|l}
\hline $N=$ & 5 & 6 & 7 & 8 & 9 & 10 &
11 & 12 \\
\hline  $t_c$ & {\tiny 1.03284}& {\tiny 1.04577 } & {\tiny
1.04302} & {\tiny 1.03743} & {\tiny 1.03199} & {\tiny
1.02739} & {\tiny 1.02353} & {\tiny 1.02036}  \\
\hline
\end{tabular}%
\caption{\emph{The critical values of t solving the PRG equation for
Fixed Parallel B.Cs}}\label{tcfixfix}%
\end{table}
\end{center}

The measured low lying excitation energies spectra corresponding
to the first case are normalized, organized in series and
presented  with the extrapolated values (obtained using the BST
algorithm \cite{BST})in table (\ref{table 3}). For the second case
see table (\ref{rrr}) for the spectrum at criticality. \\For the
fixed anti-parallel boundary conditions we observed a decreasing
and an increasing values of the pseudo-critical coupling constants
(see table (\ref{tcupdown})) which prove that in this case we
didn't reach the asymptotic regime with lengths below $N=12$. The
extrapolated values of the low lying excitation energies
corresponding to this case are given in
table.\\
In the special case of free-fixed boundary condition, the PRG
equation doesn't admit any exact solution for finite $N$. But we
remark that for the special value of $t=1$, the PRG equation
admits a minimum which converges to $0$ when $N$ \ goes to
infinity. This means that the value $t=1$ is the critical value
for infinite chain. We give in table(\ref{PRG(free_fix)}) the
corresponding values of the PRG equation for finite size chain at
$t=1$ and in table (\ref{rrr})we present the corresponding
extrapolated low lying excitation energies spectra.

\begin{center}%
\begin{table}[ht]
\begin{tabular}
[c]{lllllllll}\hline & {\tiny N=6} & {\tiny N=7} & {\tiny N=8} &
{\tiny N=9} & {\tiny N=10} & {\tiny N=11} & {\tiny N=12} & ${\tiny
N=\infty}$\\\hline {\tiny 1}$^{st}${\tiny Excitation} & {\tiny
1.00000} & {\tiny 1.00000} & {\tiny 1.00000} & {\tiny 1.00000} &
{\tiny 1.00000} & {\tiny 1.00000} & {\tiny 1.00000} & {\tiny
1.00000}\\\hline {\tiny 2\ //} & {\tiny 3.22593} & {\tiny 3.19176}
& {\tiny 3.16683} & {\tiny 3.14777} & {\tiny 3.13270} & {\tiny
3.12047} & {\tiny 3.11033} & {\tiny 3.00185}\\\hline {\tiny 3\ //}
& {\tiny 4.22593} & {\tiny 4.19176} & {\tiny 4.16683} & {\tiny
4.14777} & {\tiny 4.13270} & {\tiny 4.12047} & {\tiny 4.11033} &
{\tiny 4.00194}\\\hline {\tiny 4\ //} & {\tiny 5.20053} & {\tiny
5.19231} & {\tiny 5.18240} & {\tiny 5.17229} & {\tiny 5.16261} &
{\tiny 5.15357} & {\tiny 5.14526} & {\tiny 5.00334}\\\hline {\tiny
5\ //} & {\tiny 6.20053} & {\tiny 6.19231} & {\tiny 6.18240} &
{\tiny 6.17229} & {\tiny 6.16261} & {\tiny 6.15357} & {\tiny
6.14526} & {\tiny 6.00314}\\\hline {\tiny 6\ //} & {\tiny 6.86405}
& {\tiny 6.95879} & {\tiny 7.01551} & {\tiny 7.05058} & {\tiny
7.07267} & {\tiny 7.08664} & {\tiny 7.09540} & {\tiny
7.00484}\\\hline {\tiny 7\ //} & {\tiny 7.86405} & {\tiny 7.95879}
& {\tiny 8.01551} & {\tiny 8.05058} & {\tiny 8.07267} & {\tiny
8.08664} & {\tiny 8.09540} & {\tiny 8.00480}\\\hline {\tiny 8\ //}
& {\tiny 8.42647} & {\tiny 8.38407} & {\tiny 8.34923} & {\tiny
8.32007} & {\tiny 8.29531} & {\tiny 8.27405} & {\tiny 8.25560} &
{\tiny 8.00528}\\\hline {\tiny 9\ //} & {\tiny 8.12516} & {\tiny
8.41845} & {\tiny 8.60753} & {\tiny 8.73470} & {\tiny 8.82314} &
{\tiny 8.88630} & {\tiny 8.93238} & {\tiny 9.00406}\\\hline {\tiny
10//} & {\tiny 9.42647} & {\tiny 9.38407} & {\tiny 9.34923} &
{\tiny 9.32007} & {\tiny 9.29531} & {\tiny 9.27405} & {\tiny
9.25560} & {\tiny 9.00520}\\\hline
\end{tabular}
\caption{\emph{Measured low lying excitation energies of the Ising
quantum spin chain with free-free B.Cs}}\label{table 3}%
\end{table}
\end{center}

\begin{table}[h] \centering%
\begin{tabular}{l|l|l|l|l|l|l|l|l|l|l|l}
\hline $N=$ & 5 & 6 & 7 & 8 & 9 & 10 &
11 & 12 \\
\hline  $t_c$ & {\tiny 1.17704}& {\tiny 1.03605 } & {\tiny
0.99940} & {\tiny 0.98770} & {\tiny 0.98421} & {\tiny
0.98380} & {\tiny 0.98442} & {\tiny 0.98552}  \\
\hline
\end{tabular}%
\caption{\emph{The critical values of t solving  the PRG equation
for Fixed Anti-parallelB.Cs}}\label{tcupdown}
\end{table}
\begin{center}
\begin{table}[ht] \centering%
\begin{tabular}{l|l|l|l|l|l|l|l|l|l|l|l}
\hline $N$ & 5 & 6 & 7 & 8 & 9 & 10 &
11 & 12 \\
\hline  $PRG$ & {\tiny 0.05930}& {\tiny 0.03284 } & {\tiny
0.02012} & {\tiny 0.01325} & {\tiny 0.00921} & {\tiny
0.00667} & {\tiny 0.00500} & {\tiny 0.00384}  \\
\hline
\end{tabular}%
\caption{\emph{The value of the PRG equation at t=1 for free-fixed B.Cs}}\label%
{PRG(free_fix)}%
\end{table}%
\end{center}

\begin{center}
\begin{table}[h] \centering%
\begin{tabular}{|l|l|l|l|l|l|l|l|l|l|l|}
\hline
\multicolumn{11}{|l|}{\textbf{Fixed Parallel B.Cs}} \\
\hline & {\tiny 2.0000} & {\tiny 3.0043} & {\tiny 3.9983}& {\tiny
4.0362} & {\tiny 5.0098}&
{\tiny 5.0618}&{\tiny 6.0325}& {\tiny 6.0272}& {\tiny 6.0068} & \\
\hline \multicolumn{11}{|l|}{} \\ \hline
\multicolumn{11}{|l|}{\textbf{Free-Fixed B.Cs}}
\\ \hline & {\tiny 1.0000} & {\tiny 2.0006} & {\tiny 3.0087}& {\tiny 3.0007} & {\tiny 4.0428}&
{\tiny 4.0088}&{\tiny 5.0847}& {\tiny 5.0409}&{\tiny 5.0052} & \\
 \hline \multicolumn{11}{|l|}{}\\ \hline
 \multicolumn{11}{|l|}{\textbf{Fixed AntiParallel B.Cs}} \\
\hline & {\tiny 1.0000} & {\tiny 2.0254} & {\tiny 3.0155}& {\tiny
3.9984} & {\tiny 4.0857}&
{\tiny 5.0265}&{\tiny 5.0843}& & & \\
\hline \multicolumn{11}{|l|}{} \\ \hline
\multicolumn{11}{|l|}{\textbf{Free-Free B.Cs}} \\
\hline & {\tiny 1.0000} & {\tiny 3.0018} & {\tiny 4.0019} & {\tiny
5.0033} & {\tiny 6.0031}& {\tiny 7.0048}& {\tiny 8.0048}& {\tiny
8.0052}& {\tiny 9.0040}& {\tiny 9.0052}
\\ \hline \multicolumn{11}{|l|}{}
\\ \hline
\end{tabular}%
\caption{The low lying Energy Excitation at Criticality for the different B.Cs}\label{rrr}%
\end{table}%
\end{center}

In conclusion the analysis of the numerical measurement of the
energy excitation spectra for the Ising linear chains with
different boundary conditions confirm that these models at
criticality are in the same universality class  as those of the
boundary conformal field theories of the $(A_{2},A_{3})$ minimal model.\\
Note also that in this case all the singularities have been
considered.



\end{document}